\documentclass[12pt]{article}
\usepackage[left=2.5cm,top=2.5cm,right=2.5cm,bottom=2.5cm,nohead,nofoot]{geometry}
\usepackage{graphicx}
\usepackage{setspace}
\usepackage[square]{natbib}
\usepackage{amssymb}
\usepackage{verbatim}
\usepackage[usenames,dvipsnames]{color}

\newcommand{\imloc}{}
\newcommand{\baseloc}{../../../}

\begin{document}

\title{Censoring Outdegree Compromises Inferences of Social Network Peer Effects and Autocorrelation}
\author{Andrew C. Thomas\thanks{Visiting Assistant Professor, Department of Statistics, Carnegie Mellon University. This paper is a substantial revision of initial work done with Joe Blitzstein, who receives my thanks. I also thank Cosma Shalizi, James O'Malley, Brendan Nyhan, David Banks, those at the SAMSI Complex Networks Opening and Modelling Workshops and others for useful discussions. Part of this research was sponsored by DARPA grant 21845-1-1130102.}} 
\date{\today}

\maketitle
\thispagestyle{empty}

\begin{abstract}
I examine the consequences of modelling contagious influence in a social network with incomplete edge information, namely in the situation where each individual may name a limited number of friends, so that extra outbound ties are censored. In particular, I consider a prototypical time series configuration where a property of the ``ego'' is affected in a causal fashion by the properties of their ``alters'' at a previous time point, both in the total number of alters as well as the deviation from a central value. This is considered with three potential methods for naming one's friends: a strict upper limit on the number of declarations, a flexible limit, and an instruction where a person names a prespecified fraction of their friends. I find that one of two effects is present in the estimation of these effects: either that the size of the effect is inflated in magnitude, or that the estimators instead are centered about zero rather than related to the true effect. The degree of heterogeneity in friend count is one of the major factors into whether such an analysis can be salvaged by post-hoc adjustments.

\end{abstract}

\onehalfspacing

In any design of a social network study, there are choices to be made about the declaration of friendships between individuals so that the putative network can be studied. This often comes in the form of survey information, where respondents are asked to list all their close friends or acquaintances, or measured from behavioural observation on how the individuals interact. In the case of studies of network contagion, in which an attribute of one person is potentially passed to one or more friends (an effect also known as ``induction''), there is considerable interest in establishing a set of ``strong'' friendships that may form the backbone of the social network, through which one would potentially observe a contagious effect, such as the adoption of a technology \citep{aral2009distinguishing}, medical service or preference \citep{valente2005nmamfsdi, boulay2005sfpdpn} or physiological characteristic \citep{christakis2007solsno3y, cohen-cole2008iocsnvefoe}. 

While there are many issues regarding the perception of network ties that can distort putative models of contagion (notably the thresholding/dichotomization problem raised by \citet{thomas2010vttfl}), here I refer to a particular circumstance: the only friendships are binary in nature (as assumed by most social network models), and the respondent or ``ego'' is limited to naming a fixed number of their friends, or ``alters'' in the study in question (which I label the ``name-$k$-friends'' limitation), resulting in the omission of a number of true ties -- often the vast majority of ties.

\begin{figure}
\begin{center}
\includegraphics[width=0.5\linewidth]{\imloc 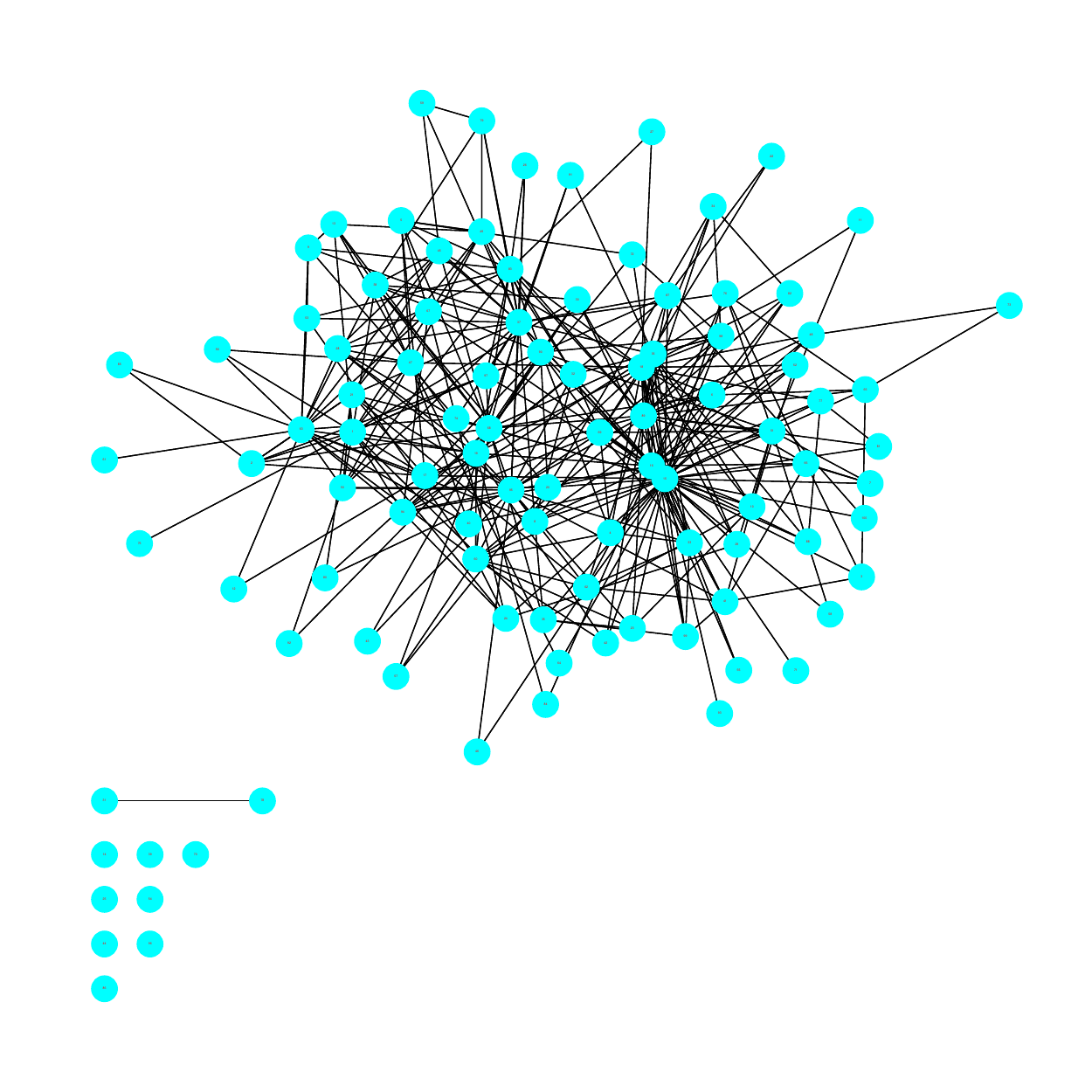}\includegraphics[width=0.5\linewidth]{\imloc 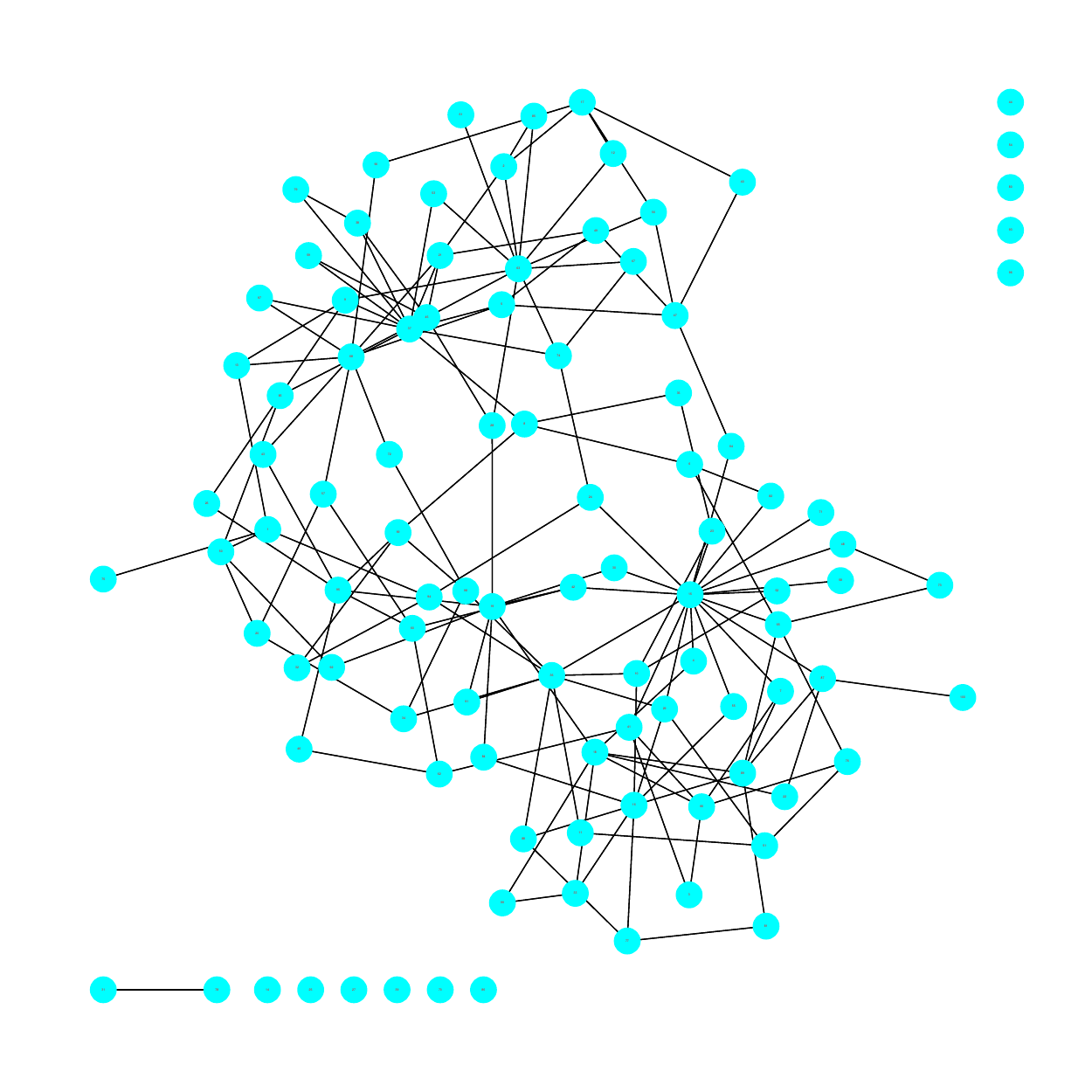}
\end{center}
\caption{The censoring process at work. Left: An uncensored 100-node binary network, with mean degree 4 and mild heterogeneity on degree. Right: The same network when each individual can name no more than two friends. One of the consequences of this censoring is the isolation of more nodes from the giant component. \label{fig:before-after}}
\end{figure}

As a result, any changes in ego behaviour between time points that were due to these ``invisible'' connections may now be attributed to properties of the self (either in terms of personal characteristics or past values of the quantity of interest) or to properties of their visible network neighbours. A direct contagion effect across an unobserved edge would instead be detected in at least one of, if not all three of, a change in the autocorrelation (self) parameter, the effect of other visible ties, or the effect of higher-degree neighbours in the system -- if the network is highly transitive (the friend of your friend is also your friend), then a person who is a true friend may appear instead as a second- or third-degree acquaintance, and models and approaches that purport to show a higher-degree pattern may instead be picking up features that are instead consequences of lower-degree actions. 

As another consequence, as seen in Figure \ref{fig:before-after}, nodes that were originally in the giant component can be disconnected if no one was able to name them, if they themselves named no one in the system. These individuals may be unwittingly driving network behaviour if contagious influence is present in the system, especially if their low popularity is connected to the trait that is undergoing contagious influence.

This treatment focuses solely on mechanisms that operate at one degree only -- that true friends of friends have no direct influence. Additionally, there are no ``false friends''; anyone who is named as a friend was truly a friend to begin with. For cases when a mechanism that censors out-degree is used, I show that there are two main consequences to this action when it comes to the estimates of coefficients in linear models: 

\begin{itemize}

\item There may be an \textbf{inflation} effect in the estimate of the coefficient size, so that the impact of any particular edgeon an individual's outcome is measured to be considerably higher than the true effect; this effect may be proportional to the fraction of ties that are observed, but can also be affected by other properties in the construction of the network.

\item There may be a \textbf{complete disruption} of the estimate of the coefficient size; rather than have an estimate that is proportional to the true underlying value, the estimate is instead due to the random (and independent) variation in the sampling mechanism, and is then centered about zero, with a variance that is an increasing function of the absolute true effect size.

\end{itemize}

\noindent Whichever effect might be observed in the analysis is a product of the naming mechanism as well as the true distribution of outgoing ties across all individuals.

I begin with examples of social network studies where out-degree censoring has been known to take place by design. I then simulate a series of binary social networks where a trait is observed on the individuals at two time steps, and make the second time step depend on both autoregressive and network-based processes. Each network naming model is then run with a series of censoring rules, and a linear model on the evolution of ego traits is run under each condition. The bias and coverage probabilities for estimators of the autocorrelated and network terms from each censored network are then compared to the ``oracle'' truth. I show that without adjusting for the censoring mechanism appropriately, the inferences made for each network effect are altered in unusual ways that will compromise many investigations.

\section{Background: ``Name Your Best Friend'' As A Network Generator\label{background}}

The National Longitudinal Study of Adolescent Health (better known as Add Health \citep{bearman1997nlsahrd}) is a study that, among other goals, aims to put health outcomes into network contexts by starting with cohorts of students within secondary schools. In order to collect information on friendship networks, participants were shown a school roster and asked to name (up to) their five best male friends and five best female friends, for a maximum of ten total friends. Romantic and sexual networks were also measured similarly, asking for ``three romantic and three non-romantic sexual partners'', for a maximum of six partners \citep{morris2004nehfsdadc}. 

For whatever reasons these constraints were put in place -- to filter out less relevant friendships and more distant sexual relationships, for example -- they were made in the design stage, well in advance of a major collection effort, and hence are beyond the ability of later investigators to expand without major modelling assumptions. Notably, 25\% of students name five friends of the same gender; 25\% name five friends of the opposite gender, and 10\% name ten total friends \citep{jackson2009dwi}.

The Framingham Heart Study (FHS) has progressed for decades as a means of tracking longitudinal medical data on a large community, but the potential of its social network information has only recently been explored. A social network was constructed in \citet{christakis2007solsno3y} after the investigators noted that a ``close friend'' was listed on the tracking information for study participants in case of lost contact, and that a great number of these friends were also in the FHS. Hence, in addition to family members, a friendship network was constructed consisting of thousands of participants.

As the designers of the heart study did not plan the social network component of future research, they did not anticipate the issue that would be caused by requesting only a single close friend for contact (though a small number of individuals listed multiple contacts regardless of the instruction). The consequences of this censoring of out-degree for friendships have not been directly addressed in the original FHS network paper and follow-up works \citep{christakis2008cdslsn, fowler2008dshlsnlao2yfhs}; in particular, whether estimations related to the relative distances between individuals would be affected by the addition of censored ties. One of the marquee claims of this research has been a universal ``three degrees of influence'' rule \citep{christakis2009csposnahtsol}, but if these claims rest on a network where a third-degree acquaintance is in reality a first-degree friend, there is considerable reason to doubt its universality.

\section{Simulation Models}

I demonstrate the impact of network structure, both real and censored, on the evolution of a trait in a networked system. There are a large number of possibilities for models of this form. I Using the notation for observables:

\begin{itemize}

\item $Y_{0i}$ and $Y_{1i}$ represent the observed trait on ego $i$ at times $t=\{0,1\}$ respectively. $\overline{Y_0}$ is the mean outcome over all individuals at time $0$.

\item $W_{ij}$ is a directional network tie, which in most applications is whether $i$ considers $j$ to be a friend and is therefore a conduit for the spread of a trait. $D_i = \sum_j W_{ij}$ is the outdegree for individual $i$, and $\overline{D}$ is the mean outdegree for all individuals.

\end{itemize}

There are a number of possible mechanisms for temporal influence on the trait in question. 

\begin{itemize}

\item \textbf{Autocorrelation}, such that the previous time point's trait will influence the present. The simplest form of this would be 

\[ Y_{1i} = \mu + \gamma Y_{0i} + \epsilon_i, \]

\noindent so that the effect is moderated by the previous time value. The effect can also be with respect to a different central point $c$, so that instead the model is $Y_{1i} = \mu' + \gamma (Y_{0i}-c) + \epsilon_i$, where $\mu = \mu' + \gamma c$. 

\item \textbf{Peer influence}, wherein a trait of an individual $j$ affects that of individual $i$ at a later time, so long as $i$ named $j$ as a friend ($W_{ij}=1$). The equation would then be 

\[ Y_{1i} = \mu + \beta \sum_j W_{ij}Y_{0j} + \epsilon_i, \]

\noindent though the effect will be more difficult to identify. If there is a different pivot point $d$, then the form is $Y_{1i} = \mu' + \beta \sum_j W_{ij}(Y_{0j}-d) + \epsilon_i'$, so that $\mu' = \mu + \beta \overline{D}$ and $\epsilon_i' = \epsilon_i + \beta(D_i - \overline{D})$. This, at least, has a diagnosis for the problem if the total peer contagion vector $WY_0$ is correlated with the trait vector $Y_0$, since this correlation can be detected after fitting.

Instead of the same pivot point for each individual, it may be that the influence is proportional to the difference in value between the alter and ego; this is essentially a ``drive toward homophily'' so that connected individuals move their values closer to each other. If this is the case, then the equation is

\[ Y_{1i} = \mu + \beta \sum_j W_{ij}(Y_{0j}-Y_{0i}) + \epsilon_i, \]

\noindent and takes additional adjustment: namely, if the model fit is $Y_{1i} = \mu' + \beta \sum_j W_{ij}Y_{0j} + \epsilon_i'$, then the adjustment becomes $\mu' = \mu + \beta \overline{DY_0}$ and $\epsilon_i' = \epsilon_i + \beta (D_iY_{0i} - \overline{DY_0})$.

\item \textbf{Peer count/outdegree effect}, so that the more friends a person has, the more their trait will change in an interval of time. In the most basic form,

\[ Y_{1i} = \mu + \delta D_{i} + \epsilon_i. \]

\noindent Again, the effect can also be with respect to a different central point $g$, so that instead the model is $Y_{1i} = \mu' + \delta (D_{i}-g) + \epsilon_i$, where $\mu = \mu' + \delta g$. 

\end{itemize}

These effects can become connected, once more than one of these terms is introduced simultaneously; as a consequence, if the specification of one term is changed, it can lead to terms other than the intercept $\mu$ to be biased. Consider the case with both peer effects in place,

\[ Y_{1i} = \mu + \beta \sum_j W_{ij}(Y_{0j}-d) + \delta D_{i} + \epsilon_i; \]

\noindent for the sake of demonstration, we identify equilibrium points on the peer influence term only. Suppose the true pivot on the contagion term was $d$ but the model chosen by the investigator chose a pivot of zero. Then the true generative equation would take the form

\[ Y_{1i} = \mu + \beta \sum_j W_{ij}Y_{0j}  - \beta (D_{i}-d) + \delta D_{i} + \epsilon_i, \]

\noindent affecting the estimate of $\mu$ (as $\mu - \beta d$) and $\delta$ (as $\delta-\beta$). Note that omitting the $D_i$ term from this equation would correspond to fixing $\delta = \beta$ in the equation, likely causing estimation error on $\beta$ if the two are different and there is any observed correlation between the terms in $WY_0$ and $Y_0$.

\subsection{Modelling the Process On The Network}

For the purpose of this investigation, I use the general evolution model of the form

\[ Y_{1i} = \mu + \gamma (Y_{0i}-\overline{Y_0}) + \beta \sum_j W_{ij} \left(Y_{0j}-\overline{Y_0}\right) + \delta \left(D_i - \overline{D} \right)  + \varepsilon_{1i}. \]

If this model faithfully represents the actual mechanism at work on the network, then the consequences of censoring network ties will be apparent in the biases of estimating the intercept $\mu$, the autocorrelation $\gamma$, the network contagion $\beta$ and the outdegree coefficient $\delta$. Essentially, the censoring mechanism takes the true sociomatrix $W$ and produces a new sociomatrix $X$, the condition being that the declared outdegree of each ego, $\sum_j X_{ij}$, is bounded above at a constant value.

Since the linear model framework represents the geometry of the covariate space, I consider two factors: how the generative mechanism of the true network relates to the prior trait value $Y_0$, and how the censoring mechanism relates to the prior trait value given the existence of the true network.

\subsection{A Network Simulation Model}

Because the geometric relationship of edge selection to trait value is primarily of interest, I construct our network according to the following formula:

\begin{itemize}

\item Determine the prior trait value $Y_0$ for all nodes in the system. For the sake of this simulation, these will be independent draws from a standard $N(0,1)$ distribution.

\item Generate a term for the gregariousness of an ego, $\alpha_i$\footnote{This follows the notation of \citet{holland1981efp} and elaborated on in \citet{thomas2010mshmfrd}.}, that is naturally heterogeneous, so that some egos seek more friendships than others. For this simulation, these will be independent draws from a normal distribution with pre-selected variance $\sigma^2_h$.\footnote{I do not induce an additional term for the differential ``popularity'' of an individual, or their tendency to attract friendships.}

\item Generate a term for homophily $h$ with respect to the trait of interest: if two individuals have similar values of the prior trait, they will be more likely to be connected if $h>0$, less likely if $h<0$ (heterophily), and indifferent if $h=0$ (isophily).

\item Choose coefficients $r_{in}$ and $r_{out}$ for the raw covariances between the pairs $(Y_{0j},W_{ij})$ and $(Y_{0i}, W_{ij})$ -- the dependence on the initial trait value and the inbound/outbound propensitis for the existence of an edge respectively. Choose the values to respect the upper bound $r_{in}^2 + r_{out}^2 < 1$.

\item Define and set a baseline continuous edge value,

\[ Z_{ij} = N(\alpha_i + r_{in}Y_{0j} + r_{out}Y_{0i} - h|Y_{0i}-Y_{0j}|, 1-(r_{in}^2 + r_{out}^2)), \] 

\noindent that will be the measure against which a binary edge is created. Create a binary network by selecting a threshold value $\omega$, and define $W_{ij} = \mathbb{I}(Z_{ij} \geq \omega)$; for the sake of this demonstration, choose $\omega$ to fix the density of arcs in the underlying graph. Ensure that there are no ``self-edges'' in the system by setting $W_{ii}=0$ for all $i$.

\end{itemize}

This model is a primitive version of the generative mechanism proposed in \citet{thomas2010mshmfrd}, but is still sufficient to explore the geometry of the network with respect to the prior trait value.

For this study, 50,000 networks were simulated with 100 or 200 nodes with varying values of each parameter. I standardize this so that each network has a mean outdegree of 10, though similar results exist for networks of different mean degree. In recognizing that many of these terms may or may not exist in a simulated network, any given parameter parameter may be set to zero in half the simulations so as to induce a variety of possible networks and evolutions.


The method of censoring outdegree is just as important to the analysis as the network construction and contagion models. I consider several different prototypical mechanisms for network censoring, the consequences of each I explore in detail.

Given that $W$ is the true friendship matrix, let $X^{(k)}$ be the friendship matrix retrieved due to the censoring/naming process. Just as the construction of the whole uncensored network can depend on the value of the trait, the naming process (conditional on these ties) can also depend on the absolute value of the traits of the contacts, the difference in value between contacts (for heterophilous or homophilous naming), or it can be completely independent of them.

The apparent equation to estimate will now be of the form

\[ Y_{1i} = \mu + \gamma (Y_{0i}-\overline{Y_0}) + \beta \sum_j X^{(k)}_{ij} \left(Y_{0j}-\overline{Y_0}\right) + \delta \left(\sum_j X^{(k)}_{ij} - \overline{\sum_j X^{(k)}_{ij}}\right)  + \varepsilon_{1i}, \]

\noindent due to the lack of information on the entire network. The consequences of each naming mechanism will then alter the geometry of $D_i$, $WY_0$ and $Y_0$ respectively; it is the consequence of these changes in geometry that I detail in the next several sections.

\section{Mechanism 1: Hard Upper Limits -- the Standard ``Name $k$ Friends'')}

\begin{figure}
\begin{center}
\includegraphics[width=\linewidth]{\imloc 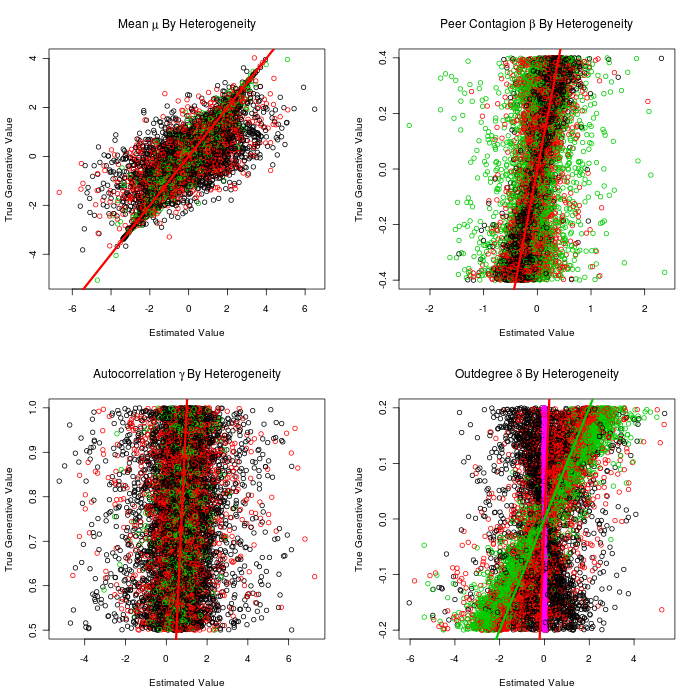}
\end{center}
\caption{Scatterplots for the estimated values of $\mu$, $\beta$, $\gamma$ and $\delta$ across all simulations for the hard upper limit of ``name one friend''. Colors refer to the degree of heterogeneity present in the system (the pink vertical refers to nonidentifiable $\delta$ values.) The red lines are along the diagonal where the estimated values would equal the true generative values; the green line represents the (expected) ``inflation'' of the effect size of $\delta$ by a factor of 10. \label{fig:strict-over}}
\end{figure}

The total number of friendships is strictly limited to be no more than $k$, and each person is required to name as many friends as they believe they have up to that limit. If each participant has at least $k$ friends, there will be an immediate difficulty: the ``number of friends'' term, $\delta \left(\sum_j X^{(k)}_{ij} - \overline{\sum_j X^{(k)}_{ij}}\right)$, reduces the covariates to a zero vector; this makes the effect of the number of friends non-identifiable.

The main burden of identification will be on two sources: the ungregarious (people with fewer than $k$ named friends), and rule-breakers (people who do name more than the appropriate number of friends.) In the case of ``name one friend'' (or $k=1$), this may lead to a vastly disproportionate degree of influence, since there are likely to be far fewer loners and rule-breakers. If the resulting mean out-degree is close to one, the few zeroes and rule-breakers will carry much of the weight in the regression, and will hypothetically add considerable bias estimates on the estimates of the effect of ``friend count''.

Figure \ref{fig:strict-over} shows the estimates for each of the model parameters for this naming scheme, as divided by the heterogeneity on outdegree in the uncensored case. There are several features that are immediately evident for each of the parameters; the outdegree effect $\delta$ is summarized in Section \ref{strict-het-sucks}. What is most obvious in the other plots is that higher heterogeneity makes the estimate for the peer contagion effect $\beta$ have higher variance, but {\em reduces} the variance on the autocorrelation $\gamma$.

\subsection{Low Heterogeneity on Degree Breaks Estimations in Strict Naming Schemes\label{strict-het-sucks}}

\begin{figure}
\begin{center}
\includegraphics[width=\linewidth]{\imloc 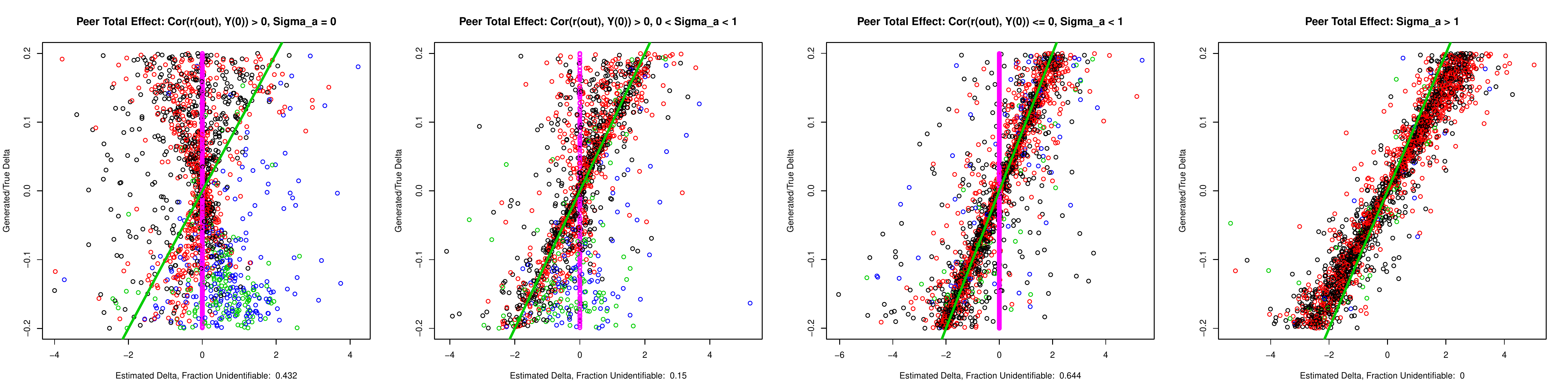}
\end{center}
\caption{Consequences of the strict name-one-friend procedure on the estimate of $\delta$, the friend-count effect, under various generative conditions. Blue and green points are cases when the estimate for the autocorrelation effect, $\widehat{\gamma}$, is negative (which it never truly is); red and black points are cases when it is positive. Pink points represent cases when $\delta$ is unidentifiable. Green lines represent a slope of 10, which is the ``expected'' inflation for the removal of 9 of 10 friendship ties. \textbf{Leftmost:} zero additional heterogeneity with positive selection on the trait of interest creates the ``cone'' effect: any potentially identifiable $\delta$ effects appear to come from a distribution centered at zero with standard deviation proportional to the true effect. \textbf{Second from the left:} once heterogeneity increases ($0<\sigma_{\alpha}<1$), the inflated effect begins to appear, with many more unidentifiable scenarios. \textbf{Second from the right:} with zero or negative correlation between in-degree and minimal heterogeneity, the ``cone'' has disappeared, so that only the zeroes and inflation are present. \textbf{Rightmost:} as heterogeneity increases, the number of zero-friend individuals increases, leaving only the inflationary case; however, the expected inflation factor begins to rise. \label{strict-delta}}
\end{figure}

There appear to be three significant families of results that appear in cases when the level of heterogeneity is varied, as seen in Figure \ref{strict-delta}:

\begin{enumerate}

\item Everyone has named a single friend, and thus the effect of relative outdegree is unidentifiable.

\item One or two people have named no friends, and therefore hold most of the power relative to the group; their personal outcomes then drive the estimate of $\delta$ and tend negative, though are still distributed about zero.

\item Several people have named no friends, and the effect is better balanced between individuals. The perceived effect size is then an ``inflation'' of the true effect size. This inflation is typically close to the ratio of total friends to named friends; this ratio increases as the number of censored friendships increases.

\end{enumerate}

The conditions that create each of these situations will vary with the generative parameters. Figure \ref{strict-delta} shows how these conditions vary with two parameters: the initial heterogeneity in outdegree between individuals, and the degree to which an individual's gregariousness varies with their own value of $Y_0$. For the sake of visualization, I remove those simulations where the true value of $\gamma$ is zero (as the sum of trials at zero is identical to those not equal to zero, and the results are similar without the benefit of a spread on the y-axis).

When the average degree is far higher than the censoring level, and heterogeneity is quite low, there will be very few if any people who have named zero friends, meaning that the effect is either unidentifiable or highly inflated. In particular, those cases in which an effect was identified, but meaningless, correspond to cases with low initial heterogeneity and a positive correlation between gregariousness and their prior $Y_0$. This suggests that those individuals with zero friends will also have low values of $Y_0$ (affecting the estimate of $\gamma$) and those with at least one friend will tend to find those with high $Y_0$ (affecting the estimate of $\beta$).

The top-left panel of Figure \ref{strict-delta} shows this in terms of colors: models with estimated $(\gamma < 0, \beta < 0$, in blue, are leading to artificially higher values of $\delta$, even when there is no true effect of network size on the outcome value. Models with estimated $(\gamma < 0, \beta > 0$, in green, appear in the majority where the true $\delta$ is negative, but the estimated value is positive.

It is worth noting that these effects virtually disappear in the case where heterogeneity is high; the estimates of $\gamma$ return to being nearly exclusively positive (as seen in the bottom right panel of Figure \ref{strict-delta}) as the estimates of $\delta$ revert to the ``inflation'' mode.

\subsection{Heterogeneity Distorts The Contagion Effect $\beta$, and Homophily on The Transmitted Attribute Biases the Autocorrelation $\gamma$\label{strict-beta-gamma}}

Homophily on the observed attribute -- the notion that two individuals are more likely to have a connection between them  if they have similar values of the attribute in question -- is a factor believed to contribute to a great deal of confounding in social network studies. For this particular section, I consider how a network with homophilous (or heterophilous) characteristics will affect linear model inferences under the action of a strict name-one-friend scheme.

\begin{figure}
\begin{center}
\includegraphics[width=\linewidth]{\imloc 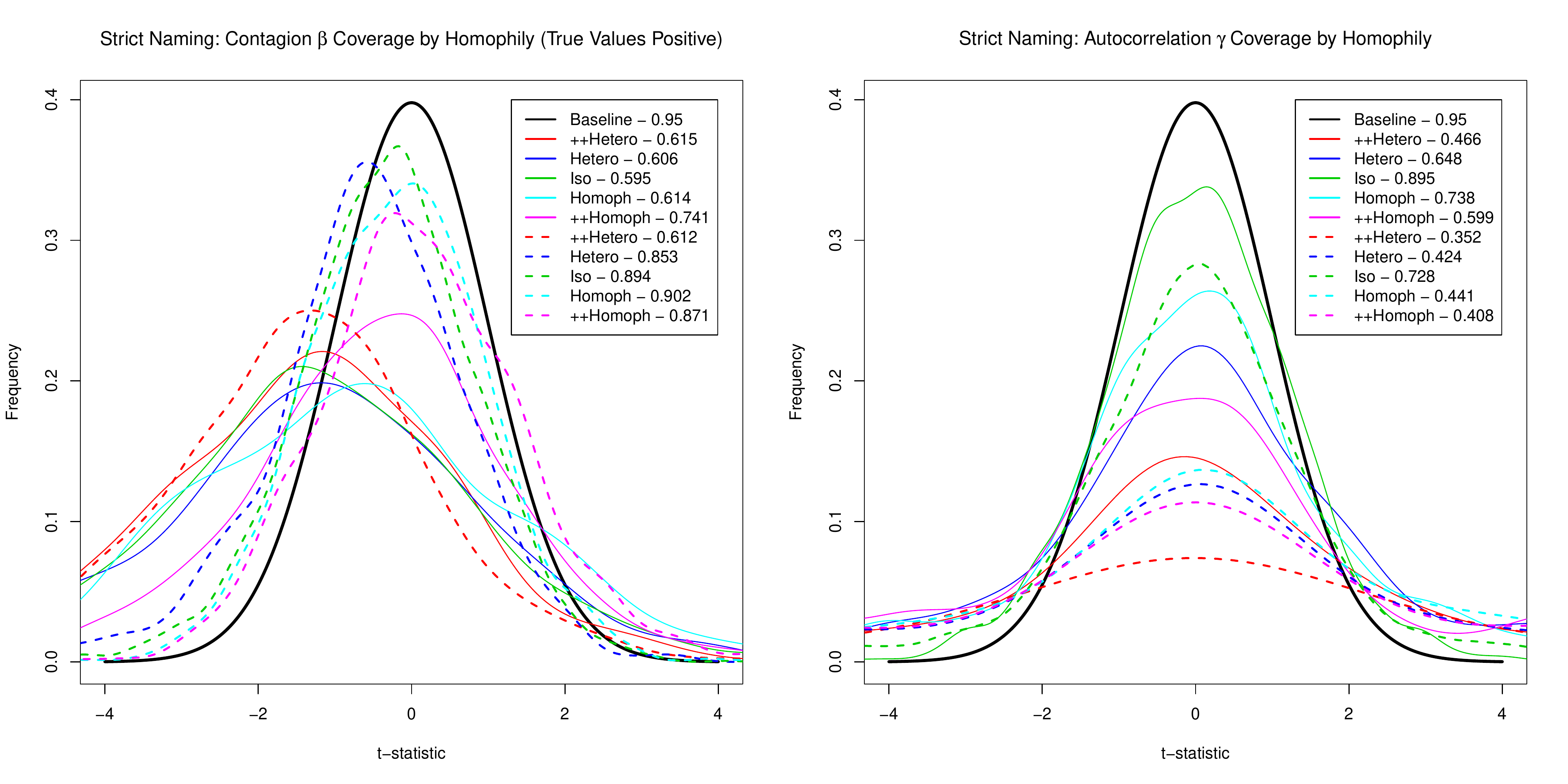}
\end{center}
\caption{Coverage properties of estimators for the contagion $\beta$ and autocorrelation $\gamma$. The solid black line is the baseline from the standard $t_{95}$ distribution; other solid lines are for high-heterogeneity cases, dashed for low-heterogeneity. Blue/cyan lines represent minimal heterophily and homophily respectively; red and pink represent strong heterophily and homophily; green lines are for isophily on the attribute. Numbers in the legend represent measured coverage probabilities for an intended 95\% confidence interval. For the contagion $\beta$, measured for positive true values, there appears to be a minimal effect of homophily on the coverage properties. If there is low heterogeneity on popularity, the estimates appear to be biased downwards, but closer inspection of Figure \ref{fig:strict-over} shows that this is in fact a disruption effect, with the interval centered about zero. For the autocorrelation effect, increased homophily or heterophily and lower heterogeneity cause coverage to be biased; high heterogeneity and zero homophily is the one case to have coverage as advertised.\label{strict-beta-gamma-homoph}}
\end{figure}

Since heterogeneity has already been shown to be the most devastating effect for the network friend count effect $\delta$, there is little point in assessing the impact of homophily on that area. Instead I examine its impact on the contagion $\beta$ and autocorrelation $\gamma$, moderating it with the impact of heterogeneity. Figure \ref{strict-beta-gamma-homoph} shows the coverage properties of the estimators in terms of a density plot of the t-statistic, $(\widehat{\beta} - \beta_{true})/\sigma_{\widehat{\beta}}$ or $(\widehat{\gamma} - \gamma_{true})/\gamma_{\widehat{\beta}}$; an estimator with the correct coverage probability should line up with the theoretical distribution of the statistic.

Heterogeneity appears to be the driving force for the estimation of the contagion $\beta$. There does not appear to be an effect of homophily or heterophily on the coverage probability, with the exception of high heterophily with low heterogeneity; this is likely related to the outdegree-dependence issue. 

Both heterogeneity and homophily/heterophily are associated with the coverage properties of the autocorrelation term $\gamma$. Coverage increases with additional heterogeneity, the opposite result. As well, additional dependence on the differences between prior characteristics decreases coverage, whether or not that dependence is positive or negative; this suggests that these friendship selections add collinearity to the autocorrelation term by establishing friendships that are similar to, or wildly different from, a person's own characteristics, and that the censoring mechanism obscures these friendship impacts by making them appear similar to the autocorrelation term.

\section{Mechanism 2: Flexible Limits -- ``Name About $k$ Friends''}

For each individual in the study, the maximum number of declared friendships is a random variable with expected value $k$; for the sake of exposition, this will be a Poisson random variable with parameter $k$, though other mechanisms are possible and produce similar results\footnote{A binomial scheme was also investigated, and produced virtually identical plots and results.}. This will induce heterogeneity in the friendship count for each person, but this heterogeneity will be driven by the naming mechanism rather than the true friendship count unless there are large numbers of low-friend individuals. This section also assumes that there is no relationship between the number of friends a person names and any other properties the person may have. 

It is important to recognize the difference between the intentional application of this naming scheme (which may prove to be more difficult) and its accidental application -- when responders violate their instructions to name (up to) a fixed number of friends. The apparent positive benefit to the researcher, in the form of additional variability on the dimension of interest, may be an illusion caused by the random variation of the naming.

\subsection{Balancing Between Naming Mechanism Noise and Inflation Effect in the Outdegree Effect $\delta$}

\begin{figure}
\begin{center}
\includegraphics[width=\linewidth]{\imloc 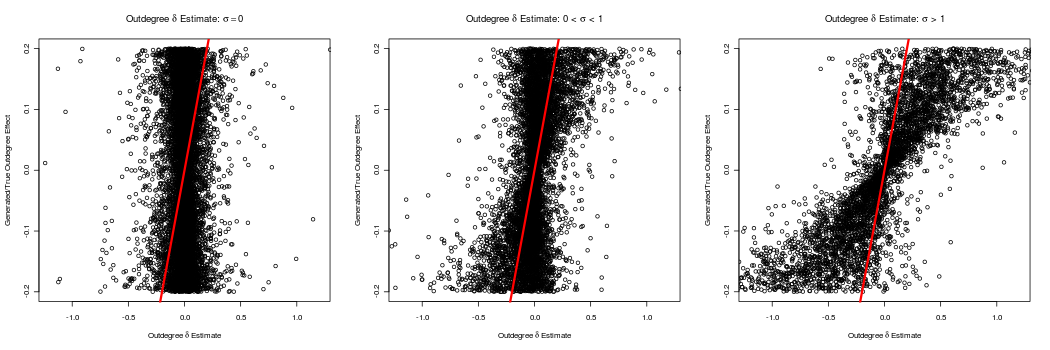}
\end{center}
\caption{The measured effect of $\delta$ as a function of heterogeneity under the flexible naming scheme. Left: with zero heterogeneity, the measured effect is centered about zero, with a mild dependence in the standard deviation as a function of the true value. As heterogeneity increases (middle, right), the zero-centered cone is replaced by a linear trend that grows past the one-to-one line, with the standard deviation increasing slowly with the absolute value of the true effect.  \label{poisson-delta-het}}
\end{figure}

Figure \ref{poisson-delta-het} shows the effect of increasing heterogeneity on the effect estimate for $\delta$. While there were three modes of effect in the strict naming case of Section \ref{strict-het-sucks}, the unidentifiability has been removed, leaving only a centered cone and a linear trend as the typical patterns of the measured effect against the truth.

In this case, low heterogeneity is wholly associated with the cone pattern -- the measured effect of $\delta$ is centered around zero, rather than the true generative value, and the width of the effect size increases with the absolute value of the true delta. This is consistent with the notion that the generative process adds signal to the system, but the naming process essentially randomizes its direction. 

As heterogeneity increases, the number of individuals with zero or one true friends increases, and the naming process will now more closely reflect the friendship counts of reality, but still distorting the signal to a degree. Notably, the inflation factor is considerably less than ten.

\subsection{Heterogeneity (Again) Distorts Estimates of The Contagion $\beta$; Homophily (Again) Affects The Autocorrelation $\gamma$\label{poisson-beta-gamma}}

\begin{figure}
\begin{center}
\includegraphics[width=\linewidth]{\imloc 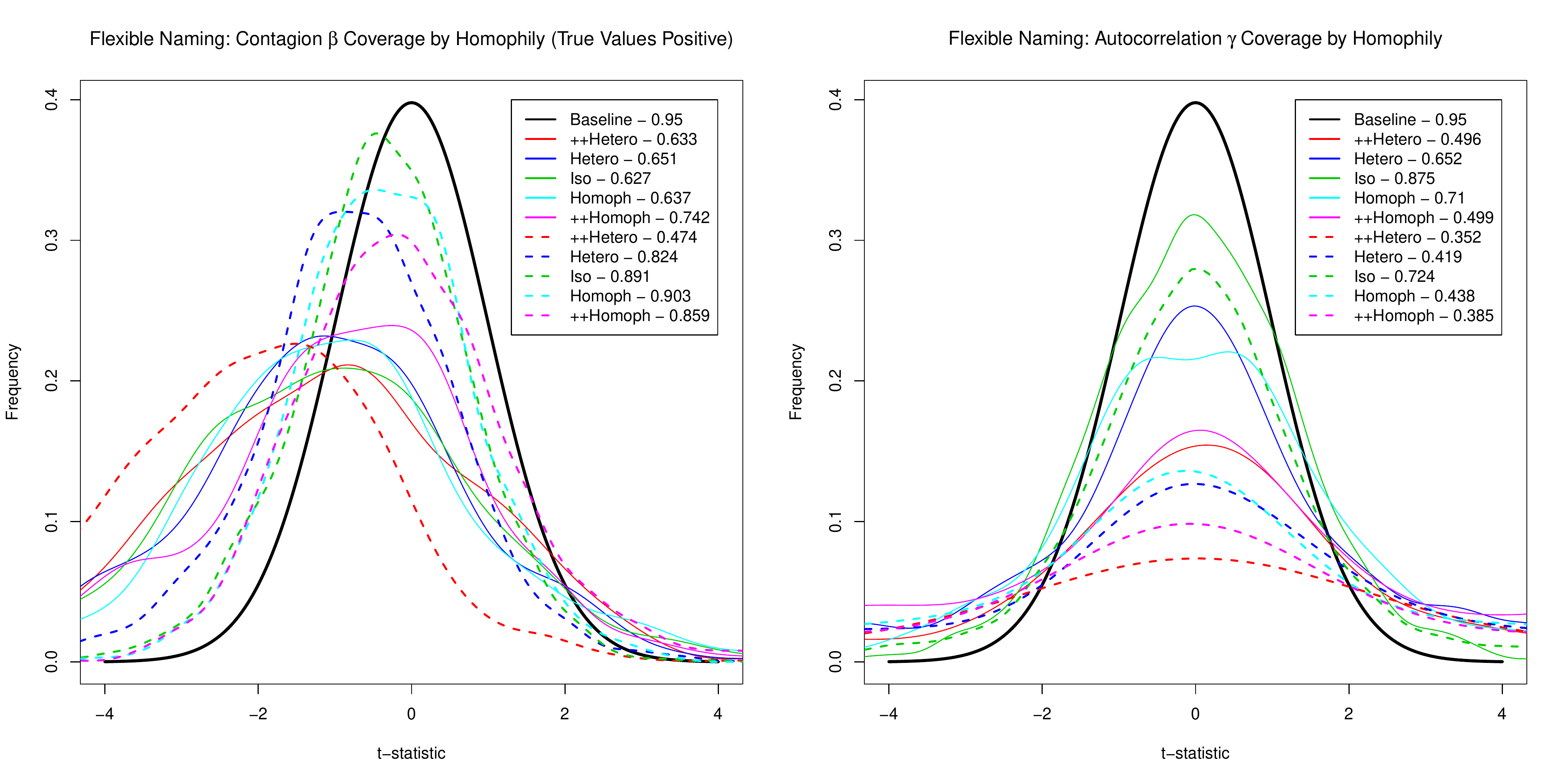}
\end{center}
\caption{Coverage properties of estimators for the network deviation term $\beta$ and autocorrelation term $\gamma$ under an approximate naming scheme. Nothing substantive has changed for the coverage probabilities of each type from the strict maximum scenario show in Figure \ref{strict-beta-gamma-homoph}. \label{poisson-beta-gamma-homoph}}
\end{figure}

The conclusions of the previous section have not changed with the implementation of this new naming scheme. Figure \ref{poisson-beta-gamma-homoph} shows the coverage properties of the estimators for $\beta$ and $\gamma$ and shows the same patterns that were shown in Section \ref{strict-beta-gamma}.

\section{Mechanism 3: Fractional Limits -- ``Name A Proportion of Your Friends'', Averaging to $k$ Per Respondent}

Since incomplete naming in many network contexts is difficult or expensive, the temptation to resort to a sampling technique is inevitable. Since varying levels of heterogeneity on outdegree have been shown to cause difficulties in cases where the naming bound is the same on everyone, it is worth investigating whether a proportional sampling method would be a better choice. First, the respondent is asked for the total number of friends they have\footnote{The definition of ``friend'' may prove to be different between people unless a standard methodology is applied in the questionnaire; see \citet{zheng2006hmp, mccormick2010hmpdykeepns} for examples on estimating network size.}; second, they are asked to name a certain fraction of those friends by name for the study under consideration. Such a method would theoretically preserve the relative gregariousness of the respondent in the total response, and hypothetically preserve the scale of the friendship count effect when incorporated into a model; if for each individual $i$ the relationship is approximated as

\[ \left(\sum_j W_{ij} - \overline{\sum_j W_{ij}}\right) \approx \frac{\overline{\sum_j W_{ij}}}{\overline{\sum_j X^{(k)}_{ij}}} \left(\sum_j X^{(k)}_{ij} - \overline{\sum_j X^{(k)}_{ij}}\right),  \]

\noindent then the relative effect of friendship count would be preserved in the average friend ratio $\frac{\overline{\sum_j W_{ij}}}{\overline{\sum_j X^{(k)}_{ij}}}$. Let the estimate for the friendship count effect using the partial network be $\widehat{\delta}$; an immediate adjustment to obtain a ``deflated'' estimate could then be $\widehat{\delta^*} = \widehat{\delta}\frac{\overline{\sum_j X^{(k)}_{ij}}}{\overline{\sum_j W_{ij}}}$, essentially dividing the inflated estimate by the fraction of friendships maintained in the sampling method.

Indeed, sub-network counts are being used in situations where the total network is difficult, if not impossible, to estimate \citep{zheng2006hmp, mccormick2010hmpdykeepns}, and such a sampling method may be sufficient to preserve the total effect of outbound friendships.

I will show that accounting for this level of heterogeneity in the sampling scheme does make several things clearer -- for example, if heterogeneity is the dominant factor in determining network structure, then correcting for the mean effect size when estimating $\delta$ is possible (though coverage probabilities will be overestimated) -- but that it complicates other aspects of the analysis.

\subsection{As Heterogeneity Rises, Mean Estimates of The Outdegree Coefficient $\delta$ Can Be Adjusted}

\begin{figure}
\begin{center}
\includegraphics[width=\linewidth]{\imloc 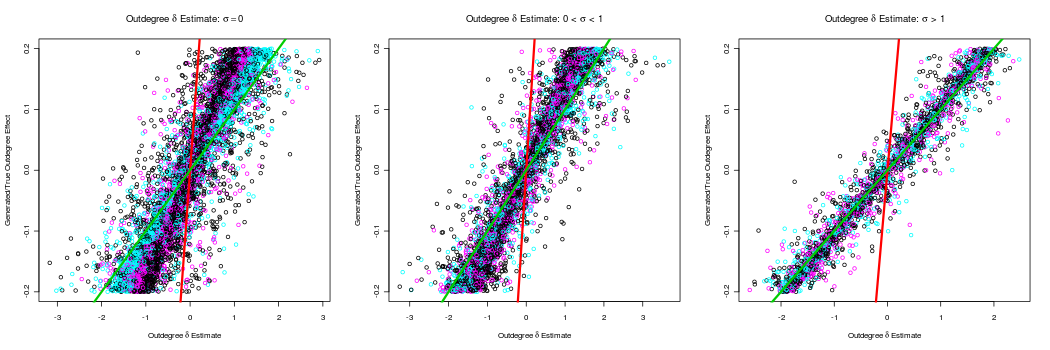}
\end{center}
\caption{Estimates of $\delta$ against true generative values under the fractional naming scheme. Black points have no generated homophily or heterophily on the attribute in question; pink points have homophily on the outcome attribute; cyan points have heterophily on the outcome attribute. In the left panel, there is no heterogeneity on the outdegree of individuals, and inflation increases when moving from isophily to homophily to heterophily. Middle, as a little heterogeneity is introduced, inflation increases though the distinction in homophily/heterophily decreases. Right, with much more heterogeneity, the effect has nearly reached the expected inflation point (here, a factor of 10) and the distinction between homophily, heterophily and isophily has disappeared.\label{fractional-delta-het}}
\end{figure}

Figure \ref{fractional-delta-het} shows the impact of various levels of heterogeneity on the estimates of delta. At the low end of heterogeneity, there is little distinction between the outdegrees of each of the individuals in the network, and the ``friend count'' effect is comparatively smaller to begin with; the rounding caused by the fractional censoring mechanism is large compared with the differences between individuals, so that the loss in resolution will diminish the inflation effect, though not eliminate it.

As the heterogeneity increases, there is far more distinction between individuals on friend count than before, and the effect becomes a greater contributor to the total variance in the outcome. The ``rounding error'' decreases, and the situation more closely approaches the complete inflation effect: for preserving one-tenth of each person's friend count, the measured effect increases tenfold. All together, this would suggest that the effect would be preserved only in cases where the possible magnitude of the signal was strong to begin with, unencumbered by the stochastic variation in the sampling mechanism and where relative ratios of friend counts can be preserved in the operation.

\subsection{Systematic Adjustments to Counter Excess Measured Contagion $\beta$}

\begin{figure}
\begin{center}
\includegraphics[width=\linewidth]{\imloc 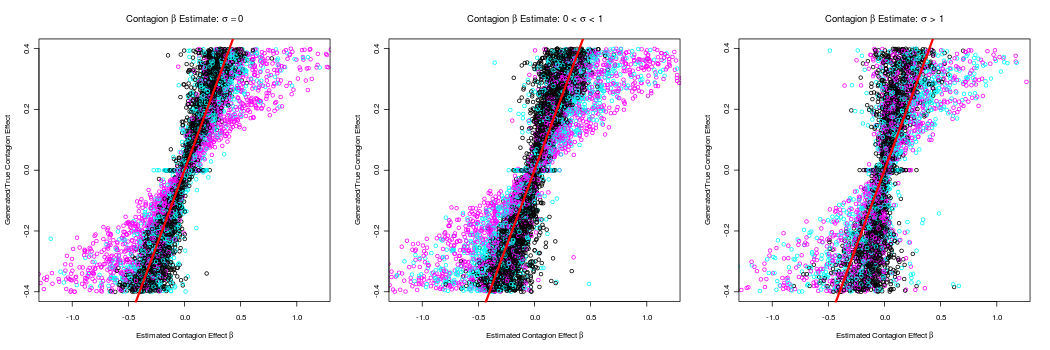}
\end{center}
\caption{Estimates of $\beta$ against true generative values under the fractional naming scheme. Black points are isophilic on the attribute in question; pink points have homophily on the outcome attribute; cyan points have heterophily on the outcome attribute. In the left panel, there is no heterogeneity on the outdegree of individuals, and inflation increases when moving from isophily to homophily to heterophily. Middle, as a little heterogeneity is introduced, inflation increases though the distinction in homophily/heterophily decreases. Right, with much more heterogeneity, much of the distinction between homophily, heterophily and isophily has disappeared.\label{fractional-beta-het}}
\end{figure}

Figure \ref{fractional-beta-het} shows the impact of various levels of heterogeneity on the estimates of the contagion effect $\beta$. In the case where there is no heterogeneity but significant homophily, the estimates of $\beta$ are inflated; as heterogeneity increases, the degree of inflation decreases. The isophilic cases also decrease in magnitude as heterogeneity increases, to the point of {\em deflation}; heterophilic cases tend to be a compromise between the homophilic and isophilic cases. In all profiles, the standard deviation of the estimates are proportional to the value of the contagion effect $\beta$ used in generating the network. 

As opposed to the outdegree effect $\delta$, these results do not immediately give rise to a corrective prescription for estimating the contagion effect $\beta$. The case of minimal heterogeneity yield a range of transformed estimates, largely inflationary in nature; as heterogeneity increases, the effect estimates are reduced in magnitude towards the truth. Homophily clearly produces an extra inflationary effect on the size of the contagion. This distinction diminishes as heterogeneity increases, yet is still present. Any prescription for correction would likely involve the estimation of the uncensored network and its propensities for homophilic selection; heterogeneity in this case may be directly estimable with accurate representations of the outdegree of each individual.

\section{Conclusions}

This investigation into the consequences of censoring of social network ties is focused on single steps: one step away in the social network, one step in time. Network processes that exist on a greater scale, both in space and time, will be affected at each scale by the omission of ties. The extreme cases under example, a tenfold reduction in the number of named ties, are given to demonstrate the phenomena that may result in the analysis of the system under a linear model.

The sampling schemes proposed fall under two categories: a constant maximum outdegree, or a measured outdegree proportional to each individual's total. The former is more likely to have naturally occurred by many naming schemes, including the studies mentioned in Section \ref{background}; the latter may prove to be a workable solution as it may naturally preserve much of the underlying geometry, but is largely a hypothetical implementation at this point and is presented for demonstration purposes as much as a proposed method of compromise between large sampling costs and losses of information.

\subsection{Consequences on Existing Studies}

As existing studies have varying levels of censoring on their outdegree, it remains to be seen how the censoring of outdegree will work in each case. The Add Health study appears to have a minimal impact at this level, with at least 75\% of respondents having named fewer friends in a category than the upper limit (assuming that the naming mechanism did not affect the naming of friends below the limit) and with the likelihood that those friends that reach the limit would not go far beyond it if the option were given. 

The Framingham Social Network, on the other hand, has considerable questions left to be answered about its naming structure, since the true distribution of friends cannot be easily assessed, especially since almost every respondent was shown to name at least one friend (even if said friend was not also in the Framingham study.) The real hope for recovery, in this case, is in the violators who named in excess of one friend -- at least one respondent named six people at one time -- though the extent of these violations is unknown. Given that the degree of censoring is likely to respected by the vast majority of respondents, the ability to reconstruct their hypothetical outdegree, let alone the complete network, may prove to introduce more error to the estimations than simply leaving them be.

This analysis used two differently specified network effects: first, the notion that simply having more friends will create an effect, and second, that a friend who is above the average level of the characteristic will have an influence on raising that characteristic. This is one interpretation of two dimensions of network effects, and many studies may have other interpretations of these dimensions. The binary traits under investigation are typically pursued only through a single indicator, whereas the inclusion of a separate friend count variable may prove to be a useful inclusion to separate its impact from the overall balance effect; whether or not the effect can be shown to be significant from zero, it may prove to reduce confounding.

\bibliographystyle{\baseloc ims}
\bibliography{\baseloc actbib}
\end{document}